\documentclass[12pt]{article}
\usepackage{amsmath,amssymb,amscd,cite}
\newtheorem{thm}{Theorem}[section]
\newtheorem{prop}[thm]{Proposition}
\newtheorem{lem}[thm]{Lemma}
\newtheorem{cor}[thm]{Corollary}
\newtheorem{defi}[thm]{Definition}
\newcommand{\pf}{{\bf Proof. \ }}
\newcommand{\qed}{\hfill $\Box$ \\}

\font\msbm=msbm10 at 12pt
\newcommand{\Z}{\mbox{\msbm Z}}

\newcommand{\F}{\mbox{\msbm F}}
\newtheorem{rem}[thm]{Remark}
\newtheorem{ex}[thm]{Example}

\begin{document}
\title{On Cyclic DNA Codes}
\author{Kenza Guenda and T. Aaron Gulliver}

\maketitle

\begin{abstract}
This paper considers cyclic DNA codes of arbitrary length over the ring $R=\F_2[u]/u^4-1$.
A mapping is given between the elements of $R$ and the alphabet~$\{A,C,G,T\}$
which allows the additive stem distance to be extended to this ring.
Cyclic codes over $R$ are designed such that their images under the
mapping are also cyclic or quasi-cyclic of index 2.
The additive distance and hybridization energy are functions
of the neighborhood energy.
\end{abstract}
\section{Introduction}
Deoxyribonucleic acid (DNA) contains the genetic program for the biological development of life.
DNA is formed by strands linked together and twisted in the shape of a double helix.
Each strand is a sequence of four possible nucleotides, two purines: adenine $(A)$
and guanine $(G)$, and two pyrimidines: thymine $(T)$ and cytosine $(C)$.
The ends of a DNA strand are chemically polar with $5'$ and $3'$ ends, which implies that the strands are oriented.
Hybridization, also known as base pairing, occurs when two strands bind together, forming a double strand of DNA.
The strands are linked following the Watson-Crick model, so that every $A$ is linked
with a $T$, and every $C$ with a $G$, and vice versa.
We denote the complement of $x$ as $\hat{x}$, i.e.,
$\hat{A}=T,\hat{T}=A,\hat{G}=C$ and $\hat{C}=G$. DNA strand pairing
is done in the opposite direction and the reverse order.
For instance, the Watson-Crick complementary (WCC) strand of $3'-ACTTAGA-5'$ is the strand $5'-TCTAAGT-3'$.
Nucleotide pairing is based on hydrogen bonds, with a pair $A-T$ forming two bonds, a pair
$G-C$ forming three bonds, and any other pair is called a mismatch because it does not form a bond.

The combinatorial properties of DNA sequences can be used to tackle
computationally difficult problems. For example,
Adleman~\cite{adleman} solved an instance of a hard (NP-complete)
computational problem, namely the directed traveling salesman
problem on a graph with seven nodes. Adleman et al.~\cite{adleman2}
used the WCC approach to break the data encryption standard (DES).
In addition, Lipton~\cite{lipton} used DNA strands to solve the
satisfiability (SAT) problem. Further, Ouyang et al.~\cite{ouyang}
presented a DNA solution to the maximum clique problem.
Since there are $4^n$ possibly single DNA strands of length $n$ which can quickly and cheaply be synthesized,
Mansuripur et al.~\cite{mansuripur} showed that DNA codewords can be used for ultra high density data storage.
Other applications exploit DNA hybridization~\cite{shoemaker}.

Software such as AMBER or CHARMM exist which can provide an accurate representation of the DNA molecule.
However, these methods are computational demanding and have a time scale on the order of $\mu s$.
This creates difficulties as many biological and other processes have time scales on the order of $ms$.
Further, these packages do not allow study of the DNA hybridization of strands (duplex formation from single strands).
This is a significant problem, as hybridization can be used as a gate in a DNA computer.
To allow parallel operations on DNA sequences, a high hybridization energy is required.
This energy depends in a rather complex way on the number of hydrogen bonds and their arrangement in the duplex.
A duplex formed by a single strand with high $GC$ content and its
reverse complement has greater stability since this pair has a high
number of hydrogen bonds. Note that in this case there are no
mismatches in the duplex. Hence the importance of designing groups
of DNA words, called a DNA code, which satisfy the reverse
complement constraint.

Breslauer et al.~\cite{breslauer} introduced the nearest-neighbor
similarity model in order to estimate the hybridization energy of a duplex.
In this model, the energy is a sum taken over pairs of positions rather than single positions.
For example, the energy of $3'-CATG-'5$ is equal to $e(CA/GT)+e(AT/TA)+e(TG/AC)$, where $e(\;)$
is the neighborhood energy of the pairs formed by the nucleotides and their WC complements, which are called stacked pairs.
This energy has been determined by experimental methods and a
comprehensive survey of these results is given in~\cite{santalucia}.
This model can be used in the ideal case, i.e., when a single strand
hybridizes with its WC complement, which is not always the case.

Secondary structure occurs when a strand folds back onto itself forming a double strand.
Milenkovic and Kashyap~\cite{olgica} argued that when designing a DNA code, a cyclic constraint
should be added to reduce the probability of secondary structure.
Secondary structure causes codewords to become computationally inactive.
This defeats the read-back mechanism in a DNA storage system by as much
as $30 \%$ as reported by Mansuripur et al.~\cite{mansuripur}.
Milenkovic and Kashyap~\cite{olgica} used the Nussinov-Jacobson
algorithm~\cite{nussinov} to prove that the presence of a cyclic
structure reduces the complexity of testing DNA codes for secondary structure.

There have been numerous results on the design of DNA codes~\cite{abualrub,abualrub1,bahattin,gaborit,G-GDNA}.
The problem of hybridization energy has also been studied extensively~\cite{bishop,dyachkov01,dyachkov03,siap}.
More recently, D'Yachkov et al.~\cite{dyachkov09} modeled the hybridization energy for DNA strand as an
additive stem similarity using the neighborhood energy of pairs of nucleotides.
They also introduced the additive stem distance.
Bahattin and Siap~\cite{bahattin} constructed DNA codes as cyclic
reversible complement codes of odd length over the ring
\[
R=\F_2[u]/(u^4-1)=\{a+bu+cu^2+du^3\, |\, a,b,c,d \in F_2, u^4=1\}.
\]
They also studied the problem of the Hamming distance.

In this paper we construct cyclic DNA codes of arbitrary length over the ring $R=\F_2[u]/u^4-1$.
This is a finite chain ring with 16 elements.
A mapping is given between the elements of $R$ and the alphabet~$\{A,C,G,T\}$
which allows the notion of additive stem distance to be extended to this ring.
Cyclic codes are obtained over $R$ which are reversible-complement and have images under the mapping
which are also cyclic or quasi-cyclic of index 2.
They also satisfy the WCC condition, and the additive distances and hybridization energies can be determined.
Note that one can also find a one-to-one to map between the elements $\{A,C,G,T\}^2$ and the field of cardinality 16,
but codes over a ring are more suitable.
In particular, codes over rings can contain more codewords than similar codes over fields,
and they provide more flexibility in constructing codes.
Moreover, there exists more cyclic codes over rings than over fields.
The structure of repeated cyclic codes over finite chain rings is in general not known.
We use the fact that $\F_2 \subset R$ and they have the same characteristic
to find the structure of these codes.
Note that the results given hold for any finite chain ring with cardinality 16 and characteristic 2.
For example, $\F_2+u\F_2+u^2\F_2+u^3\F_2$ with $u^4=0$ is such a ring.
Typically the Hamming distance or deletion distance are used in designing DNA codes.
However, these metrics do not capture the thermodynamic properties and combinatorial structure of DNA.
We consider the additive stem distance and adapt it for use with our DNA codes.
Another reason for using the ring $R$ is that the codes can be mapped to
DNA codes of length $2n$ which contain a subcode with large $GC$-content and thus
has a high hybridization energy.

The next section presents some basic facts and preliminaries.
Section 3 introduces the additive stem-similarity model.
Then cyclic DNA codes are investigated in Section 4.
In particular, the structure of cyclic codes of even length over the ring $R$ is determined.
Our DNA codes are presented and the stem-similarity distance is extended to these codes.

\section{Preliminaries}

The ring considered here is
\[
R=\F_2[u]/(u^4-1)=\{a+bu+cu^2+du^3\, |\, a,b,c,d \in F_2, u^4=1\},
\]
which is a commutative ring with 16 elements.
It is a principal local ideal with maximal ideal $\langle u+1 \rangle$.
The ideals satisfy
\begin{equation}
\label{eq:ch} \langle 0 \rangle = \langle (u+1)^4 \rangle \subsetneq
\langle (u+1)^3 \rangle \subsetneq \langle (u+1)^2 \rangle
\subsetneq \langle (u+1) \rangle \subsetneq R.
\end{equation}
The field $\F_2$ is a subring of $R$, a fact which will be used later.

A map $\phi$ which is a one-to-one correspondence between the
elements of $R$ and the DNA nucleotide base pairs $\{A, T, C, G\}^2
$ is given in Table~\ref{tab:santalucia2}.
\begin{table}
\caption{Correspondence between the nucleotide base pairs and the elements of $R$.} \label{tab:santalucia2}
\begin{center}
\begin{tabular}{cccccccc}
\hline
$GG$&0&$AT$ &$1+u$&$GT$&1&$CT$&$1+u+u^2$ \\
$CC$&$1+u+u^2+u^3$ &$TA$&$u^2+u^3$ & $TG$&$u^2$ &$TC$&$1+u^2+u^3$ \\
$GC$&$1+u^2$&$AA$&$u+u^2$&$AC$&$1+u+u^3 $& $AG$& $u$ \\
$CG$& $u+u^3$&$TT$ & $1+u^3$&$CA$&$u+u^2+u^3$&$GA$&$u^3$  \\
\hline
\end{tabular}
\end{center}
\end{table}
A simple verification gives that for all $x\in R$, we have
\begin{equation}
\label{eq:hat} x+\hat{x}=u^3+u^2+u+1.
\end{equation}
In addition, multiplying an element $x$ of $R$ by $u^2$ reverses the DNA pair corresponding to $x$.
Further, multiplying any $x\in R$ by $u^2$ reverses the corresponding pair in $\{A,G,C,T\}^2$.
Note that other mappings can be defined between $R$ and the nucleotide pairs~\cite{bahattin}.
The mapping $\phi$ was chosen because it results in a subcode over the alphabet~$\{GC,CC,GG,CG\}$
which will have a high hybridization energy.

Since $R^n$ is an $R$ module, a linear code over $R$ of length $n$ is a submodule $\mathcal{C}$ of $R^n$.
Now let $\mathcal{A}$ any alphabet.
Then a code $\mathcal{C}$ over $\mathcal{A}$ is said to be
\textbf{cyclic} if it is invariant under a cyclic shift, i.e.,
$(x_{n-1},x_0,\ldots,x_{n-2})\in \mathcal{C}$ provided the codeword
$(x_0,x_1,\ldots,x_{n-2},x_{n-1})$ is in $\mathcal{C}$.
A code is called \textbf{quasi-cyclic} of index $l$ if for any
$(x_0,x_1,\ldots,x_{n-2},x_{n-1})\in \mathcal{C}$ we have
$(x_{n-l},x_{n-l+1},\ldots,x_{0},x_{1},x_{n-l-1}) \in \mathcal{C}$.
Note that these definitions hold regardless of whether the code is linear.
The structure of linear cyclic codes of length $n$ over $R$ when $n$ is odd has been
examined in~\cite{CS, permounth}, but the general case has not yet been investigated.

%
Let $x=x_0x_1\ldots x_{n-1}$ be a vector in $R^n$.
The reverse of $x$ is defined to be $x^r=x_{n-1}x_{n-2}\ldots x_{1}x_0$,
the complement of $x$ is $x^c=\hat{x}_0\hat{x}_1\ldots \hat{x}_{n-1}$,
and the reverse complement (also called the Watson-Crick complement) is
$x^{rc}=\hat{x}_{n-1}\hat{x}_{n-2}\ldots \hat{x}_{1} \hat{x}_{0}$.
A code $\mathcal{C}$ is said to be \textbf{reverse complement} if
for any $x\in \mathcal{C}$ we have $x^{rc} \in \mathcal{C}$.
\begin{defi}
For $\mathcal{A}=\{A,G,C,T\}$, a code $\mathcal{C}$ of length $n$
over the alphabet $\mathcal{A}$ is called reversible if the WCC of
each codeword $a\in \mathcal{A}^n$ is also in $\mathcal{C}$.
\end{defi}

\section{Additive Stem Similarity Distance}

In this section, the additive stem similarity introduced by D'yachkov and Voronina \cite{dyachkov09} is presented.
The DNA hybridization energy for strands $x$ and $y$ is an important measure
of the stability of the duplex, as it is related to the melting temperature of the duplex.
This is the temperature required to melt a duplex.
The hybridization energy of a duplex can be modeled as a function of the so-called
neighborhood energy of the nucleotides.
For a pair $a,b \in \mathcal{A}=\{A,C,G,T\}$, the neighborhood energy is given by
\[
w(a,b)=\Delta G(a,b)= \Delta H(a,b)-T\Delta S(a,b),
\]
where $\Delta H(a,b)$ and $\Delta S(a,b)$ are the
temperature-independent enthalpy and entropy, respectively.
The pairs $(a,b)\in \mathcal{A}$ are also called stacked pairs.
For example, these quantities as well as $\Delta G(a,b)$ for a
temperature of $310^ \circ$ are given in Table~\ref{tab:santalucia}.
\begin{table}
\caption{Nearest Neighbor Thermodynamic Values for Stacked Pairs~\cite{bishop}} \label{tab:santalucia}
\begin{center}
\begin{tabular}{cccc}
\hline
 Stacked pair & && \\
$5' \rightarrow 3' /3' \rightarrow 5'$& $\Delta H\, kcal/mol$&
$\Delta S\, kcal/mol$& $\Delta G_{310 ^ \circ} \, kcal/mol$  $n$ \\
\hline
$AA/TT=TT/AA$ & $-7.9$ &$-22.2$&$-1.02$ \\
$AC/TG=GT/CA$& $-8.4$&$ -22.4$ & $-1.46$ \\
$AG/TG=CT/GA$&$-7.8$& $-21.0$& $-1.29$ \\
$AT/TA$&$-7.2$&$-20.4$& $-0.88$ \\
 $CA/GT=TG/AC$&$-8.5$& $-22.7$ & $-1.46$\\
 $CC/GG=GG/GC$&$-8.0$& $-19.9$ & $-1.83$\\
 $CG/GC$&$-10.6$& $-27.2$ & $-2.17$\\
 $GA/CT=TC/AG$&$-8.2$& $-22.2$ & $-1.32$\\
 $GC/CG$&$-9.8$& $-24.4$ & $-2.24$\\
$TA/AT$&$-7.2$& $-21.3$ & $-0.60$\\
\hline
\end{tabular}
\end{center}
\end{table}
For $x=x_1,\ldots,x_n \in \mathcal{A}^n$ and $y=y_1,\ldots,y_n \in \mathcal{A}^n$, define
\[
S_w(x,y)=\sum_{i=1}^{n-1}s_i^w(x,y),
\]
where
\[
s_i^w(x,y)=\left\{
\begin{array}{ll}
 w(a,b)& \text{ if } x_i=y_i=a,x_{i+1}=y_{i+1}=b, \\
 0 & \text{ otherwise}. \\
\end{array}
\right.
\]
and $w(a,b)$ is the neighborhood energy of the pair $(a,b)\in \mathcal{A}^2$.
The quantity $S_w(x,y)$ is called the additive stem similarity between $x$ and $y$, and
it satisfies the following properties
\[
S_w(x,y)=S_w(y,x) \le S_w(x,x).
\]
The hybridization energy between $x$ and $y$ is~\cite{dyachkov09}
\[
E(x,y)=S_w(x,y^{rc}).
\]
\begin{defi}
\label{defi:stem} Let $x$ and $y$ in $\mathcal{A}^n$. Then the real
number
\[D(x,y)= S_w(x,x)-S_w(x,y)\]
is called the additive stem distance between $x$ and $y$ in
$\mathcal{A}^n.$
\end{defi}
It is clear that $D(x,x)=0$, but in general it is not symmetric and does not satisfy the triangle inequality.

\section{Cyclic DNA Codes}
Let $\mathcal{C}$ be a linear code over $R$.
$\mathcal{A}=\{A,C,G,T\}$ and $D(.,.)$ be the additive stem distance
given in Definition~\ref{defi:stem}.
Since the map $\phi$ defined in Table~\ref{tab:santalucia2} is one-to-one, the additive stem distance can be extended to the ring $R$
as follows.
For $x,y \in R$, define the additive stem distance over $R$ as
\begin{equation}
\label{eq:stem3}
 D(x,y)=D(\phi(x),\phi(y)).
 \end{equation}
Let $D=\min_{x\neq y} D(x,y)$ for $x,y \in \mathcal{C}$.
A cyclic DNA code over $R$ is then defined as follows.
\begin{defi}
\label{def:1}
A cyclic code $\mathcal{C}$ over $R$ is called an $[n,d]$ cyclic DNA code if it
satisfies the following:
\begin{itemize}
\item $\mathcal{C}$ is a cyclic code, i.e., $C$ is an ideal in $R_n=R[x]/(x^n-1)$;
\item for any codeword $x\in \mathcal{C}$, $x \neq x^{rc}$ and $x^{rc} \in \mathcal{C}$; and
\item $D(x,y) \ge d, \forall x,y \in \mathcal{C}$.
\end{itemize}
\end{defi}

Let $\mathcal{C}$ be an $[n,d]$ cyclic DNA code over $R$.
Then if $s=\max\{S_w(\phi(x), \phi(x)), x\in \mathcal{C} \}$,
from (\ref{eq:stem3}) and the definition of the additive stem distance we obtain that
\[
S_w((\phi(x),\phi(y)) \le S_w((\phi(x),\phi(x))-D((\phi(x),\phi(y)), \forall x,y \in \mathcal{C}.
\]
Therefore $ S_w(\phi(x),\phi(y)) \le s-d$ for all $x,y \in \mathcal{C}$, and thus in our context a cyclic DNA code over $R$ is a
cyclic reverse-complement code such that
\begin{equation}
E(\phi(x),\phi(y)) \le s-d,\, \forall x,y \in  \mathcal{C}.
\end{equation}
\begin{defi}
A code $\mathcal{C}$ over an alphabet $\mathcal{A}$ is called an
$(n,d)$ DNA code if it is a block code of length $n$ such that
$D(x,y) \ge d$ for all $x,y\in \mathcal{C}.$
\end{defi}
\subsection{Cyclic Codes over $R$ of Arbitrary Length}
The purpose of this section is to determine the structure of cyclic codes over the ring $R$.
Only codes of even length are considered as the structure of cyclic codes over $R$ of odd length has
previously been examined~\cite{CS,permounth}.
In the case $n$ odd it has been proven that the cyclic codes over $R$ are in fact principal ideals.
This is not true for the case $n$ even.

We begin by providing some results for codes of odd length.
\begin{lem}
A cyclic code of odd length $n$ over $R$ is an ideal defined as
\begin{equation} \label{eq:gen}
 \mathcal{C}=\langle f_0|(u+1)f_1|(u+1)^2f_2|(u+1)^3f_3\rangle
\end{equation}
such that $f_3|f_2|f_1 |f_0 |x^n-1$.
\end{lem}
Note that there exists a canonical surjective ring morphism $(-)$ given by
\begin{equation}
\label{eq:over}
\begin{split}
(-):  R[x]&  \longrightarrow \F_2[x] \\
f &\longmapsto \overline{f}=f \mod u+1
\end{split}
\end{equation}
\begin{defi}
A polynomial $f$ in $R[x]$ is called regular if $\overline{f} \neq
0$. $f$ is called primary if the ideal $\langle f \rangle$ is
primary, and $f$ is called basic irreducible if $\overline{f}$ is
irreducible in $F_2[x]$. Two polynomials $f$ and $g$ in $R[x]$ are
called coprime if
$$R[x]=\langle f \rangle +\langle g \rangle.$$
\end{defi}

\begin{lem}(\cite[Theorem XIII. 11]{Mac})
\label{XIIIMac} Let $f$ be a regular polynomial in $R[x]$. Then
$f=\alpha g_1 \ldots g_r$, where $\alpha$ is a unit and $g_1 ,
\ldots , g_r$ are regular primary coprime polynomials. Moreover,
$g_1, \ldots , g_r$ are unique in the sense that if $f=\alpha g_1
\ldots g_r=\beta h_1 \ldots h_s$, where $\alpha, \beta$ are units
and $g_i$ and $h_i$ are regular primary coprime polynomials, then
$r=s$, and after renumbering $\langle g_i \rangle= \langle h_i
\rangle,$ $1\le i \le n$.
\end{lem}

\begin{lem}
\label{lem:basic} If $f(x)\in R[x]$ is a basic irreducible
polynomial, then $f(x)$ is a primary polynomial.
\end{lem}
\pf Assume that $f(x)$ is basic irreducible and $g(x)h(x)\in \langle
f(x) \rangle$.
Then $\overline{f}(x)$ is irreducible in $K[x]$, so that
$(\overline{f}(x),\overline{g}(x))=1$ or $ \overline{f}(x)$. If
$(\overline{f}(x),\overline{g}(x))=1$ then $f$ and $g$ are also
coprime, and there exist $f_{1}$ and $g_{1}$ in $R[x]$ such that
$1=f(x)f_{1}(x)+g(x)g_{1}(x)$. Hence
$h(x)=f(x)h(x)f_{1}(x)+g(x)h(x)g_{1}(x)$. Since $g(x)h(x)\in \langle
f(x) \rangle$, it follows that $h(x)\in  \langle f(x)  \rangle$. If
$(\overline{f}(x),\overline{g}(x))=\overline{f}(x)$, then there
exist $f_1(x), g_1(x)\in R[x]$ such that
$g(x)=f(x)f_1+(u+1)^{i}g_1(x)$ for some positive integer $i<4$. Then
for $k>i$, we have $g^{k}\in \langle f(x) \rangle$, and thus $f(x)$
is a primary polynomial. \qed
\begin{rem}
\label{rem:fact} Let $m$ be an odd integer. Then from~\cite{G-G} the
polynomial $x^m-1$ factors uniquely as a product of monic basic
irreducible pairwise coprime polynomials over $R$, and there is a
one-to-one correspondence between the set of irreducible divisors in
$\F_2$. Since $\F_2$ is a subring of $R$ and the decomposition of
$x^m-1$ is unique in $R$, the polynomials $f_i$ are in $\F_2$.
\end{rem}

\begin{prop}
\label{prop:deco} If $n=m2^s$ such that $m$ is an odd integer, then
$x^n-1$ has a unique decomposition over $R$ given by
\begin{equation}
\label{eq:fact} x^n-1={g_1}^{2^s} \ldots {g_l}^{2^s},
\end{equation}
where the $g_i$ are irreducible polynomials coprime in $\F_2[x]$
which are divisors of $x^m-1$.
\end{prop}
\pf For any integer $s \ge 0$ and odd $m \ge 0$. We have that
$(x^m-1)^{2^s}=x^{m2^s}-1$ because $2| \binom{2^s}{i}$ for $1\le i
\le 2^s$. From Remark~\ref{rem:fact}, $x^m-1$ has a unique
decomposition into irreducible polynomials over $\F_2$ as follows:
$x^m-1=g_1 \ldots g_l$. We need to prove that $x^n-1=g_1^{2^s}
\ldots g_l^{2^s}$ is unique. Assume that $x^n-1=f_1^{\alpha_1}\ldots
f_r^{\alpha_l}$ is a decomposition into powers of basic irreducible
polynomials. From Lemma~\ref{lem:basic} we have that the basic
irreducible polynomials are primary, hence the power of a basic
irreducible polynomial is also a primary polynomial. Then from
Lemma~\ref{XIIIMac}, the decomposition (\ref{eq:fact}) is unique.

\begin{prop}
\label{prop:main} With the previous notation, the primary ideals of
$\mathcal{R}$ are $\langle 0 \rangle$, $\langle 1 \rangle$, $\langle
g_i^j \rangle$, $\langle g_i^j,(u+1)^t \rangle$, with $1 \le j \le
2^s$, $1\le t \le 3 $ and $1\le i \le l$.
\end{prop}
\pf  Let $\mu : R[x]\longmapsto \frac{\F_{2}[x]}{\langle
x^n-1\rangle }$ be the canonical homomorphism.
By~Lemma~\ref{prop:deco}, we have that the factorization of
$x^n-1=g_1^{2^s} \ldots g_l^{2^s}$ over $R[x]$ is the same as that
over $\F_{2}[x]$ and is unique. This gives that the kernel of $\mu$
is the ideal $\langle x^n-1,u\rangle$. Hence from~\cite[Theorem
3.9.14]{Z-S}, the primary ideals of $\mathcal{R}$ are the preimages
of the primary ideals of $\F_{2}[x]/x^n-1$. It is well
known~\cite[Theorem 3.10]{Mac2} that the primary ideals of this last
ring are the ideals $\langle g_i^j\rangle$, $1\le j\le 2^s$ and
$1\le i \le l$. Hence the primary ideals of $\mathcal{R}$ are
$\langle g_i^j, (u+1)^t \rangle$.
\qed

\begin{thm}
\label{th:gen} Let $n=m2^s$ such that $m$ is an odd integer.
 Then the cyclic codes of length $2^sm$ over $R$ are the ideals generated by $\langle
f_0|(u+1)f_1|(u+1)^2f_2|(u+1)^3f_3 \rangle$, where
$f_3|f_2|f_1|f_0|x^n-1$.
\end{thm}
\pf Let $\mathcal{C}$ be a cyclic code in $R[x]$ so that
$\mathcal{C}$ is an ideal of $\mathcal{R}$. Since $\mathcal{R}$ is
Noetherian, from the Lasker-Noether decomposition Theorem \cite[p.
209]{Z-S}, any ideal in $\mathcal{R}$ has a representation as a
product of primary ideals. From Proposition~\ref{prop:main}, we have
that the primary ideals of $\mathcal{R}$ are $\langle g_i^j, (u+1)^t
\rangle$, where $x^n-1=\prod_{l=1}^r {g_i}^{2^s}$. Hence an ideal
$I$ of $\mathcal{R}$ is of the form
\begin{equation}
\label{eq:form}  I=\prod_{l=1}^r \langle  g_i^j, (u+1)^t \rangle.
\end{equation}
Expanding the product in~(\ref{eq:form}), each ideal in $
\mathcal{R}$ is generated by
\[\langle f_0|(u+1)f_1|(u+1)^2f_2|(u+1)^{3}f_{3} \rangle,
\]
where $f_3|f_2|f_1|f_0|x^n-1$.
\qed

\subsection{The Reverse-Complement Constraint}
In this section, the reverse-complement constraint is examined for
cyclic codes of arbitrary length $n$. Denote
$(x^{n}-1)/(x-1))=\mathbb{I}(x)$. The following lemma will be used
later.
 \begin{lem}[\cite{abualrub}]
 \label{lem:selreci}
Let $f(x)$ and $g(x)$ be polynomials in $R[x]$ with $\deg f(x) \ge
\deg f(x)$. Then the following holds:
\begin{itemize}
\item[(i)] $[f(x)g(x)]^*=f(x)^*g(x)^*$;
\item[(ii)] $[f(x)+g(x)]^*=f(x)^*+x^{\deg f- \deg g }g(x)^*$.
\end{itemize}
\end{lem}
\begin{thm}
\label{lem:1} Let $\mathcal{C}$ be a reverse-complement cyclic code
over $R$. Then the following holds:
\begin{itemize}
\item[(i)] $\mathcal{C}$ contains the codeword $(1+u+u^2+u^3)\mathbb{I}(x)$.

\item[(ii)] $\mathcal{C}=\langle f_0|(u+1)f_1|(u+1)^2f_2|(u+1)^3f_3\rangle$ with  all $f_i$ self-reciprocal.
\end{itemize}
\end{thm}
\pf Part (i) is from~\cite{bahattin}.
Part (ii) in the case $n$ odd was proven in~\cite[Theorem 4.3]{bahattin}.
Since by Theorem~\ref{th:gen} the codes of even length are generated by
$\mathcal{C}=\langle f_0|(u+1)f_1|(u+1)^2f_2|(u+1)^3f_3\rangle$,
the argument in~\cite{bahattin} for $n$ odd also holds for the case $n$ even.
\qed

The proof of the following Theorem is the same as that of~\cite[Theorem 4.4]{bahattin} for odd length.
\begin{thm}
\label{th:self3} Let $\mathcal{C}$ be a cyclic code over $R$ of
length $n$. Suppose $(1+u+u^2+u^3)\mathbb{I}(x)\in \mathcal{C}$.
Then if $\mathcal{C}= \langle
f_0|(u+1)f_1|(u+1)f_1^2|(u+1)^3f_3\rangle$, where the $f_i$ are
self-reciprocal, then $\mathcal{C}$ is a reverse-complement code.
\end{thm}
\begin{cor}
\label{cor:dual} Let $\mathcal{C}$ be a cyclic code of length
$n=2^sm,$ $s \ge 0$. Then if $(1+u+u^2+u^3)\mathbb{I}(x)\in
\mathcal{C}$ and if there exists an $i$ such that
\begin{equation}
\label{eq:dual} 2^i\equiv -1 \mod m,
\end{equation}
then $\mathcal{C}$ is a reverse-complement code.
\end{cor}
\pf  Let $\mathcal{C}=\langle f_0 | (u+1 )f_1|(u+1 )^2f_2|(u+1)^3f_3
\rangle$ be a cyclic code of length $n$. The polynomials $f_i$ are
divisors of $x^n-1$ in $\F_2$. The decomposition into the product of
minimal polynomials is given by $x^n-1=\prod M_i(x)$. Each $M_i$
corresponds to a cyclotomic class $Cl(i)$. Equation~(\ref{eq:dual})
gives that $Cl(1)$ is reversible and hence all the cyclotomic
classes are reversible. Thus each minimal polynomial is
self-reciprocal, and from Lemma~\ref{lem:selreci} the polynomials
$f_i$ are self-reciprocal. Then from Theorem~\ref{th:self3},
$\mathcal{C}$ is a reverse-complement code.
%
\qed
\begin{ex}
Let $n=6$.
Then the cyclic code over $R$ with generator polynomial
$(1+u+u^2+u^3)(x^2+x+1)$ is a cyclic reversible code over $R$.
\end{ex}

\begin{cor}
\label{corThmain}
Let $\mathcal{C}$ be an $[n,d]$ cyclic DNA code over $R$.
Then $\phi(\mathcal{C})$ is a $[2n,d]$ quasi-cyclic $DNA$ code of index 2 over the alphabet $\{A,G,C,T\}$.
\end{cor}
\pf Let $\mathcal{C}$ be a cyclic DNA code of length $n$ over $R$.
Hence $\phi(\mathcal{C})$ is a set of length $2n$ over the alphabet
$\mathcal{A}$ which is quasi-cyclic of index 2.
Since $\mathcal{C}$ is a reverse-complement code, then $u^2x^{rc} \in \mathcal{C}$,
and $\phi(u^2x^{rc})$ is the WCC of $\phi(x)$.
\qed

\begin{defi}
For a code $\mathcal{C}$ over $R$, define the subcode $\mathcal{C}_{1+u^2}$ consisting
of all codewords in $\mathcal{C}$ that are a multiple of $(1+u^2)$.
\end{defi}
\begin{lem}
\label{lem12} With the previous definition we have
\[
\phi((1+u^2)R)=\{GG,CC,CG,GC\}.
\]
Further, if $\mathcal{C}=\langle f_0 | (u+1 )f_1|(u+1)^2f_2|(u+1)^3f_3 \rangle$
is a cyclic code of length $n$ over $R$, then
\[
\mathcal{C}_{1+u^2} = \langle (1+u^2)f_3(x) \rangle.
\]
\end{lem}
\pf The first part of the lemma is a simple verification.
For the second part, assume $\mathcal{C}=\langle f_0 | (u+1)f_1|(u+1)^2f_2|(u+1)^3f_3 \rangle$.
Since $f_3|f_2$ then we have $\langle (1+u^2) f_3 \rangle \subset \mathcal{C}_{1+u^2}$.

Now let $c(x)\in \mathcal{C}$ so that $c(x)=k_0(x)f_0(x)+ (u+1)k_1(x)f_1(x)+(u+1)^2k_2(x)f_2(x)+(u+1)^3k_3(x)f_3(x)$ for $k_i(x) \in \F_2[x]$.
If $c(x)$ is a multiple of $1+u^2$, then we have $x^n-1 | k_0(x)f_0(x)$ and $x^n-1 |k_0(x)f_1(x)$ and hence
$c(x)=(1+u^2)((k_2(x)f_2(x)+(1+u)(k_3(x)f_3(x))$.
Since $f_3(x)|f_2(x)|f_1(x)$, $\mathcal{C}_{1+u} \subset \langle
f_3(x)\rangle$, and therefore $\mathcal{C}_{1+u^2}= \langle (1+u^2)f_3 \rangle$.
\qed

Let $d_{1+u^2}=\min\{D(x,y), \; x,y \in \mathcal{C}_{1+u^2}\}$. Then
the following holds.
\begin{thm}
\label{th:mai44} Let $\mathcal{C}=\langle f_0 | (u+1 )f_1|(u+1
)^2f_2|(u+1)^3f_3 \rangle$ be an $[n,d]$ cyclic DNA code over $R$.
Then $\phi (C_{1+u^2})$ is a cyclic DNA code of length $n$ over the
alphabet $\{GG,CC,GC,CG\}$ such that $d_{1+u^2} \ge d$.
\end{thm}
\pf Since $\mathcal{C}_{1+u^2} \subset \mathcal{C}$, it is obvious that $d_{1+u^2} \ge d$.
From Theorem~\ref{corThmain}, we have that the image of the cyclic DNA code $\mathcal{C}$
obtained via $\phi$ is a quasi-cyclic code of length $2n$ over the alphabet $\{A,G,C,T\}$.
From Lemma~\ref{lem12} we have that $\phi((1+u^2)R)=\{GG,CC,CG,GC\}$ and
$\mathcal{C}_{1+u^2} = \langle (1+u^2)f_3(x) \rangle$.
This gives the result.
\qed

\begin{rem}
Theorem~\ref{th:mai44} is useful as it results in cyclic subcodes with large $GC$-content.
Since from Table~\ref{tab:santalucia} the stacked pair corresponding to $\{GG,CC,GC,CG\}$ has the largest neighborhood energy,
these subcodes high high hybridization energy.
\end{rem}

\begin{center}
\textbf{Acknowledgements}
\end{center}
The authors would like to thank Anne Condon for drawing their
attention to the problem of the nearest neighbor energy model.

\end{document}